\documentclass[10pt,a4paper,onecolumn]{article}
\usepackage{marginnote}
\usepackage{graphicx}
\usepackage{xcolor}
\usepackage{authblk,etoolbox}
\usepackage{titlesec}
\usepackage{calc}
\usepackage{tikz}
\usepackage{hyperref}
\hypersetup{colorlinks,breaklinks=true,
            urlcolor=[rgb]{0.0, 0.5, 1.0},
            linkcolor=[rgb]{0.0, 0.5, 1.0}}
\usepackage{caption}
\usepackage{tcolorbox}
\usepackage{amssymb,amsmath}
\usepackage{ifxetex,ifluatex}
\usepackage{seqsplit}
\usepackage{xstring}

\usepackage{float}
\let\origfigure\figure
\let\endorigfigure\endfigure
\renewenvironment{figure}[1][2] {
    \expandafter\origfigure\expandafter[H]
} {
    \endorigfigure
}

\usepackage{fixltx2e} % provides \textsubscript
\usepackage[
  backend=biber,
]{biblatex}
\bibliography{paper.bib}

\let\textttOrig=\texttt
\def\texttt#1{\expandafter\textttOrig{\seqsplit{#1}}}
\renewcommand{\seqinsert}{\ifmmode
  \allowbreak
  \else\penalty6000\hspace{0pt plus 0.02em}\fi}

\makeatletter
\let\href@Orig=\href
\def\href@Urllike#1#2{\href@Orig{#1}{\begingroup
    \def\Url@String{#2}\Url@FormatString
    \endgroup}}
\def\href@Notdoi#1#2{\def\tempa{#1}\def\tempb{#2}%
  \ifx\tempa\tempb\relax\href@Urllike{#1}{#2}\else
  \href@Orig{#1}{#2}\fi}
\def\href#1#2{%
  \IfBeginWith{#1}{https://doi.org}%
  {\href@Urllike{#1}{#2}}{\href@Notdoi{#1}{#2}}}
\makeatother

\newlength{\cslhangindent}
\newlength{\csllabelwidth}
\newenvironment{CSLReferences}[2] % #1 hanging-ident, #2 entry spacing
 {% don't indent paragraphs
  \setlength{\parindent}{0pt}
  \ifodd #1 \everypar{\setlength{\hangindent}{\cslhangindent}}\ignorespaces\fi
  \ifnum #2 > 0
  \setlength{\parskip}{#2\baselineskip}
  \fi
 }%
 {}
\usepackage{calc}
\newcommand{\CSLBlock}[1]{#1\hfill\break}
\newcommand{\CSLLeftMargin}[1]{\parbox[t]{\csllabelwidth}{#1}}
\newcommand{\CSLRightInline}[1]{\parbox[t]{\linewidth - \csllabelwidth}{#1}}
\newcommand{\CSLIndent}[1]{\hspace{\cslhangindent}#1}

\usepackage[top=3.5cm, bottom=3cm, right=1.5cm, left=1.0cm,
            headheight=2.2cm, reversemp, includemp, marginparwidth=4.5cm]{geometry}

\renewcommand\familydefault{\sfdefault}

\renewcommand{\bibfont}{\small \sffamily}
\renewcommand{\captionfont}{\small\sffamily}
\renewcommand{\captionlabelfont}{\bfseries}

\titleformat{\section}
  {\normalfont\sffamily\Large\bfseries}
  {}{0pt}{}
\titleformat{\subsection}
  {\normalfont\sffamily\large\bfseries}
  {}{0pt}{}
\titleformat{\subsubsection}
  {\normalfont\sffamily\bfseries}
  {}{0pt}{}
\titleformat*{\paragraph}
  {\sffamily\normalsize}

\usepackage{fancyhdr}
\renewcommand{\headrulewidth}{0pt}
\renewcommand{\footrulewidth}{0.25pt}

\makeatletter
\let\ps@plain\ps@fancy
\fancyheadoffset[L]{4.5cm}
\fancyfootoffset[L]{4.5cm}

\definecolor{linky}{rgb}{0.0, 0.5, 1.0}

\newtcolorbox{repobox}
   {colback=red, colframe=red!75!black,
     boxrule=0.5pt, arc=2pt, left=6pt, right=6pt, top=3pt, bottom=3pt}

\newcommand{\ExternalLink}{%
   \tikz[x=1.2ex, y=1.2ex, baseline=-0.05ex]{%
       \begin{scope}[x=1ex, y=1ex]
           \clip (-0.1,-0.1)
               --++ (-0, 1.2)
               --++ (0.6, 0)
               --++ (0, -0.6)
               --++ (0.6, 0)
               --++ (0, -1);
           \path[draw,
               line width = 0.5,
               rounded corners=0.5]
               (0,0) rectangle (1,1);
       \end{scope}
       \path[draw, line width = 0.5] (0.5, 0.5)
           -- (1, 1);
       \path[draw, line width = 0.5] (0.6, 1)
           -- (1, 1) -- (1, 0.6);
       }
   }

\patchcmd{\@maketitle}{center}{flushleft}{}{}
\patchcmd{\@maketitle}{center}{flushleft}{}{}
\patchcmd{\@maketitle}{\LARGE}{\LARGE\sffamily}{}{}
\def\maketitle{{%
  
  \AB@maketitle}}
\makeatletter
\renewcommand\AB@affilsepx{ \protect\Affilfont}
\renewcommand\AB@affilnote[1]{{\bfseries #1}\hspace{3pt}}
\renewcommand{\affil}[2][]%
   {\newaffiltrue\let\AB@blk@and\AB@pand
      \if\relax#1\relax\def\AB@note{\AB@thenote}\else\def\AB@note{#1}%
        \setcounter{Maxaffil}{0}\fi
        \begingroup
        \let\href=\href@Orig
        \let\texttt=\textttOrig
        \let\protect\@unexpandable@protect
        \def\thanks{\protect\thanks}\def\footnote{\protect\footnote}%
        \@temptokena=\expandafter{\AB@authors}%
        {\def\\{\protect\\\protect\Affilfont}\xdef\AB@temp{#2}}%
         \xdef\AB@authors{\the\@temptokena\AB@las\AB@au@str
         \protect\\[\affilsep]\protect\Affilfont\AB@temp}%
         \gdef\AB@las{}\gdef\AB@au@str{}%
        {\def\\{, \ignorespaces}\xdef\AB@temp{#2}}%
        \@temptokena=\expandafter{\AB@affillist}%
        \xdef\AB@affillist{\the\@temptokena \AB@affilsep
          \AB@affilnote{\AB@note}\protect\Affilfont\AB@temp}%
      \endgroup
       \let\AB@affilsep\AB@affilsepx
}
\makeatother
\renewcommand\Authfont{\sffamily\bfseries}
\renewcommand\Affilfont{\sffamily\small\mdseries}
\ifnum 0\ifxetex 1\fi\ifluatex 1\fi=0 % if pdftex
  \usepackage[T1]{fontenc}
  \usepackage[utf8]{inputenc}

\else % if luatex or xelatex
  \ifxetex
    \usepackage{mathspec}
    \usepackage{fontspec}

  \else
    \usepackage{fontspec}
  \fi
  \defaultfontfeatures{Ligatures=TeX,Scale=MatchLowercase}

\fi
\IfFileExists{upquote.sty}{\usepackage{upquote}}{}
\IfFileExists{microtype.sty}{%
\usepackage{microtype}
\UseMicrotypeSet[protrusion]{basicmath} % disable protrusion for tt fonts
}{}

\usepackage{hyperref}
\hypersetup{unicode=true,
            pdftitle={MSG: A software package for interpolating stellar spectra in pre-calculated grids},
            pdfborder={0 0 0},
            breaklinks=true}
\let\addcontentslineOrig=\addcontentsline
\def\addcontentsline#1#2#3{\bgroup
  \let\texttt=\textttOrig\addcontentslineOrig{#1}{#2}{#3}\egroup}
\let\markbothOrig\markboth
\def\markboth#1#2{\bgroup
  \let\texttt=\textttOrig\markbothOrig{#1}{#2}\egroup}
\let\markrightOrig\markright
\def\markright#1{\bgroup
  \let\texttt=\textttOrig\markrightOrig{#1}\egroup}

\usepackage{graphicx,grffile}
\makeatletter
\def\maxwidth{\ifdim\Gin@nat@width>\linewidth\linewidth\else\Gin@nat@width\fi}
\def\maxheight{\ifdim\Gin@nat@height>\textheight\textheight\else\Gin@nat@height\fi}
\makeatother
\setkeys{Gin}{width=\maxwidth,height=\maxheight,keepaspectratio}
\IfFileExists{parskip.sty}{%
\usepackage{parskip}
}{% else
\setlength{\parindent}{0pt}
\setlength{\parskip}{6pt plus 2pt minus 1pt}
}
\providecommand{\tightlist}{%
  \setlength{\itemsep}{0pt}\setlength{\parskip}{0pt}}
\ifx\paragraph\undefined\else
\let\oldparagraph\paragraph
\renewcommand{\paragraph}[1]{\oldparagraph{#1}\mbox{}}
\fi
\ifx\subparagraph\undefined\else
\let\oldsubparagraph\subparagraph
\renewcommand{\subparagraph}[1]{\oldsubparagraph{#1}\mbox{}}
\fi

\title{MSG: A software package for interpolating stellar spectra in
pre-calculated grids}

        \author[1]{Rich Townsend}
          \author[1]{Aaron Lopez}
    
      \affil[1]{Department of Astronomy, University of
Wisconsin-Madison, USA}
  \date{\vspace{-7ex}}

\begin{document}
\maketitle

\marginpar{

  \begin{flushleft}
  %\hrule
  \sffamily\small

  {\bfseries DOI:} \href{https://doi.org/10.21105/joss.04602}{\color{linky}{10.21105/joss.04602}}

  \vspace{2mm}

  {\bfseries Software}
  \begin{itemize}
    \setlength\itemsep{0em}
    \item \href{https://github.com/openjournals/joss-reviews/issues/4602}{\color{linky}{Review}} \ExternalLink
    \item \href{https://github.com/rhdtownsend/msg}{\color{linky}{Repository}} \ExternalLink
    \item \href{https://doi.org/10.5281/zenodo.7559319}{\color{linky}{Archive}} \ExternalLink
  \end{itemize}

  \vspace{2mm}

  \par\noindent\hrulefill\par

  \vspace{2mm}

  {\bfseries Editor:} \href{}{} \ExternalLink \\
  \vspace{1mm}
    \vspace{2mm}

  {\bfseries Submitted:} 5 June 2022\\
  {\bfseries Published:} 29 January 2023

  \vspace{2mm}
  {\bfseries License}\\
  Authors of papers retain copyright and release the work under a Creative Commons Attribution 4.0 International License (\href{http://creativecommons.org/licenses/by/4.0/}{\color{linky}{CC BY 4.0}}).

  \end{flushleft}
}

\hypertarget{summary}{%
\section{Summary}\label{summary}}

While the spectrum of the light emitted by a star can be calculated by
simulating the flow of radiation through each layer of the star's
atmosphere, this process is computationally expensive. Therefore, it is
often far more efficient to pre-calculate spectra over a grid of
photospheric parameters, and then interpolate within this grid.
\texttt{MSG} (short for Multidimensional Spectral Grids) is a software
package that implements this interpolation capability.

\hypertarget{statement-of-need}{%
\section{Statement of Need}\label{statement-of-need}}

There are a wide variety of stellar spectral grids published in the
astronomical literature --- examples include Lanz and Hubeny (2003),
Lanz and Hubeny (2007), Kirby (2011), de Laverny et al. (2012), Husser
et al. (2013), Allende Prieto et al. (2018), Chiavassa et al. (2018) and
Zsargó et al. (2020). However, the ecosystem of software packages that
offer users the ability to interpolate in these grids is much more
limited:

\begin{itemize}
\item
  FERRE (Allende-Prieto and Apogee Team 2015) supports piecewise-cubic
  interpolation in an arbitrary number of photospheric parameters, but
  is restricted to grids with rectilinear boundaries. Moreover, as a
  monolithic executable it is not well suited to modular embedding
  within other projects.
\item
  Starfish (Czekala et al. 2015) offers a Python API supporting
  piecewise-linear interpolation in an arbitrary number of photospheric
  parameters (see Mészáros and Allende Prieto 2013 for a discussion of
  the limitations of linear schemes).
\item
  stsynphot (STScI Development Team 2020) also offers a Python API
  supporting piecewise-linear interpolation, but is restricted to three
  photospheric parameters and a hard-coded selection of grids.
\end{itemize}

The limitations of these packages stem in part from their purpose: each
has a broader focus than spectral interpolation alone. Guided by the
Unix philosophy of `make each program do one thing well' (McIlroy,
Pinson, and Tague 1978), this motivates us to develop MSG.

\hypertarget{capabilities}{%
\section{Capabilities}\label{capabilities}}

MSG is implemented as a software library with Python, Fortran 2008 and C
bindings. These APIs each provide routines for interpolating specific
intensity and flux spectra. They are underpinned by OpenMP-parallelized
Fortran code that performs energy-conservative interpolation in
wavelength \(\lambda\), parametric interpolation in direction cosine
\(\mu\) using limb-darkening laws, and \(C^{1}\)-continuous cubic
tensor-product interpolation in an arbitrary number of photospheric
parameters (effective temperature \(T_{\mathrm{eff}}\), surface gravity
\(g\), metallicity {[}Fe/H{]}, etc.). Although the topology of grid
points must remain Cartesian, their distribution along each separate
dimension need not be uniform. Attempts to interpolate in regions with
missing data (e.g., ragged grid boundaries and/or holes) are signalled
gracefully via exceptions (Python) or returned status codes (Fortran and
C).

To minimize disk space requirements, MSG grids are stored in HDF5
container files with a flexible and extensible schema. Tools are
provided that can create these files from existing grids in other
formats. Rather than reading an entire grid into memory during program
start-up (which is slow and may not even be possible, given that some
grids can be hundreds of gigabytes in size), MSG loads data into a cache
only when needed; and once the cache size reaches a user-specified
limit, data are evicted using a least-recently-used algorithm.

In addition to specificintensity and flux, MSG can evaluate associated
quantities such as moments of the radiation field. It can also convolve
spectra on-the-fly with filter/instrument response functions, to provide
corresponding photometric colors. Therefore, it is a straightforward and
complete solution to synthesizing observables (spectra, colors, etc.)
for stellar models, and serves as an ideal seasoning to add flavor to
stellar astrophysics research.

\hypertarget{acknowledgments}{%
\section{Acknowledgments}\label{acknowledgments}}

We are grateful to the late Keith Smith for laying the original
foundations for MSG, and likewise acknowledge support from NSF grant
ACI-1663696 and NASA grant 80NSSC20K0515.

\hypertarget{references}{%
\section*{References}\label{references}}
\addcontentsline{toc}{section}{References}

\hypertarget{refs}{}
\begin{CSLReferences}{1}{0}
\leavevmode\vadjust pre{\hypertarget{ref-Allende-Prieto:2018}{}}%
Allende Prieto, C., L. Koesterke, I. Hubeny, M. A. Bautista, P. S.
Barklem, and S. N. Nahar. 2018. {``{A collection of model stellar
spectra for spectral types B to early-M}.''} \emph{Astronomy \&
Astrophysics} 618 (October): A25.
\url{https://doi.org/10.1051/0004-6361/201732484}.

\leavevmode\vadjust pre{\hypertarget{ref-Allende-Prieto:2015}{}}%
Allende-Prieto, Carlos, and Apogee Team. 2015. {``{FERRE: A Code for
Spectroscopic Analysis}.''} In \emph{American Astronomical Society
Meeting Abstracts}, 225:422.07.

\leavevmode\vadjust pre{\hypertarget{ref-Chiavassa:2018}{}}%
Chiavassa, A., L. Casagrande, R. Collet, Z. Magic, L. Bigot, F.
Thévenin, and M. Asplund. 2018. {``{The STAGGER-grid: A grid of 3D
stellar atmosphere models. V. Synthetic stellar spectra and broad-band
photometry}.''} \emph{Astronomy \& Astrophysics} 611 (March): A11.
\url{https://doi.org/10.1051/0004-6361/201732147}.

\leavevmode\vadjust pre{\hypertarget{ref-Czekala:2015}{}}%
Czekala, Ian, Sean M. Andrews, Kaisey S. Mandel, David W. Hogg, and
Gregory M. Green. 2015. {``{Constructing a Flexible Likelihood Function
for Spectroscopic Inference}''} 812 (2): 128.
\url{https://doi.org/10.1088/0004-637X/812/2/128}.

\leavevmode\vadjust pre{\hypertarget{ref-de-Laverny:2012}{}}%
de Laverny, P., A. Recio-Blanco, C. C. Worley, and B. Plez. 2012.
{``{The AMBRE project: A new synthetic grid of high-resolution FGKM
stellar spectra}.''} \emph{Astronomy \& Astrophysics} 544 (August):
A126. \url{https://doi.org/10.1051/0004-6361/201219330}.

\leavevmode\vadjust pre{\hypertarget{ref-Husser:2013}{}}%
Husser, T. -O., S. Wende-von Berg, S. Dreizler, D. Homeier, A. Reiners,
T. Barman, and P. H. Hauschildt. 2013. {``{A new extensive library of
PHOENIX stellar atmospheres and synthetic spectra}.''} \emph{Astronomy
\& Astrophysics} 553 (May): A6.
\url{https://doi.org/10.1051/0004-6361/201219058}.

\leavevmode\vadjust pre{\hypertarget{ref-Kirby:2011}{}}%
Kirby, Evan N. 2011. {``{Grids of ATLAS9 Model Atmospheres and MOOG
Synthetic Spectra}.''} \emph{Proceedings of the Astronomical Society of
the Pacific} 123 (903): 531. \url{https://doi.org/10.1086/660019}.

\leavevmode\vadjust pre{\hypertarget{ref-Lanz:2003}{}}%
Lanz, T., and I. Hubeny. 2003. {``{A Grid of Non-LTE Line-blanketed
Model Atmospheres of O-Type Stars}.''} \emph{Astrophysical Journal
Supplement} 146 (June): 417--41. \url{https://doi.org/10.1086/374373}.

\leavevmode\vadjust pre{\hypertarget{ref-Lanz:2007}{}}%
---------. 2007. {``{A Grid of NLTE Line-blanketed Model Atmospheres of
Early B-Type Stars}.''} \emph{Astrophysical Journal Supplement} 169
(March): 83--104. \url{https://doi.org/10.1086/511270}.

\leavevmode\vadjust pre{\hypertarget{ref-McIlroy:1978}{}}%
McIlroy, M. D., E. N. Pinson, and B. A. Tague. 1978. {``UNIX
Time-Sharing System: Forward.''} \emph{Bell System Technical Journal} 57
(6): 1899.

\leavevmode\vadjust pre{\hypertarget{ref-Meszaros:2013}{}}%
Mészáros, Sz., and C. Allende Prieto. 2013. {``{On the interpolation of
model atmospheres and high-resolution synthetic stellar spectra}.''}
\emph{Monthly Notices of the Royal Astronomical Society} 430 (4):
3285--91. \url{https://doi.org/10.1093/mnras/stt130}.

\leavevmode\vadjust pre{\hypertarget{ref-STScI:2020}{}}%
STScI Development Team. 2020. {``{stsynphot: synphot for HST and
JWST}.''} Astrophysics Source Code Library, record ascl:2010.003.

\leavevmode\vadjust pre{\hypertarget{ref-Zsargo:2020}{}}%
Zsargó, J., C. R. Fierro-Santillán, J. Klapp, A. Arrieta, L. Arias, J.
M. Valencia, L. Di G. Sigalotti, M. Hareter, and R. E. Puebla. 2020.
{``{Creating and using large grids of precalculated model atmospheres
for a rapid analysis of stellar spectra}.''} \emph{Astronomy \&
Astrophysics} 643 (November): A88.
\url{https://doi.org/10.1051/0004-6361/202038066}.

\end{CSLReferences}

\end{document}